# Spin-layer Locked Gapless States in Gated Bilayer Graphene

W. Jaskólski[*]

Institute of Physics, Faculty of Physics,

Astronomy and Informatics, Nicolaus Copernicus University,

Grudziadzka 5, 87-100 Torun, Poland

A. Ayuela

Donostia International Physics Center (DIPC),

Paseo Manuel Lardizabal 4, 20018 Donostia-San Sebastián, Spain

Centro de Física de Materiales, CFM-MPC CSIC-UPV/EHU,

Paseo Manuel Lardizabal 5, 20018 Donostia-San Sebastián, Spain and

Departamento de Física de Materiales, Facultad de Químicas,

UPV-EHU, 20018 San Sebastián, Spain

(Dated: June 18, 2019)

## Abstract

Gated bilayer graphene exhibits spin-degenerate gapless states with a topological character localized at stacking domain walls. These states allow for one-dimensional currents along the domain walls. We herein demonstrate that these topologically protected currents are spin-polarized and locked in a single layer when bilayer graphene contains stacking domain walls decorated with magnetic defects. The magnetic defects, which we model as $\pi$-vacancies, perturb the topological states but also lift their spin degeneracy. One gapless state survives the perturbation of these defects, and its spin polarization is largely localized in one layer. The spin-polarized current in the topological state flows in a single layer, and this finding suggests the possibility of effectively exploiting these states in spintronic applications.

[*]Electronic address: `wj@fizyka.umk.pl`

## I. INTRODUCTION

Bilayer graphene (BLG) has attracted a great deal of attention because it can exist with a variety of stacking arrangements with intriguing electronic properties [1, 2]. Of particular note is the revival of interest in twisted layers of graphene due to their superconductivity [3]. In the search for graphene-based systems for nanoelectronic applications [4-10], Bernal-stacked BLG has considerable promise as it possesses a tunable energy gap under an applied gate voltage [11-14]. Furthermore, gated BLG containing stacking domain walls (SDWs) shows conducting gapless states, which are topologically protected and well defined in terms of the valley index [15-18]. These states continue to attract attention for their potential applications in novel devices and as testing grounds for topological phases in 2D materials [19, 20].

The existence of topological states has been reported in experiments on gapped BLG [21, 22]. Topological states appear as a pair of states connecting the valence and conduction band continua in the energy gap of gated BLG [15, 17, 18, 20]. They are induced by SDWs, i.e., when the order of the sublattice changes from AB to BA. Experimentally, a SDW can be realized by stretching, corrugating, folding, or introducing defects in one of the layers [18, 21, 23, 24]. In a bilayer, topological states allow for transport along the domain wall [18]. These states are robust and topologically protected because the valley Chern number changes across the SDW [16, 20]. Although these states were first described in topological terms, atomistic modelling is required to study their physico-chemical properties in experimental realizations [25]. Atomistic calculations extend beyond topological considerations and predict the survival of topological states in cases with atomic-scale defects present in SDWs [24].

The conducting states that arise at the one-dimensional boundary between two regions of gated BLG with different stacking orders are topologically valley protected [16]. The gapless states are thus locked in their momentum and characterized by pseudospin due to the sublattice symmetry. Additionally, the states are spin degenerate, and it may be necessary to remove the spin degeneracy for future applications, e.g., spintronic devices. According to Lieb's theorem for π-vacancies in 2D systems, the spin degeneracy can be removed by a magnetic signal driven by a local imbalance between the A and B sublattices [26]. Graphene π-vacancies can be created by a large number of experimental methods using

Co substituents, radicals, H-doping, and even fluorination [27, 28]. We investigate how the spin degeneracy of gapless states is removed by magnetic defects near SDWs in BLG.

In this work, we consider the vacancy defects close to the SDW in one of the layers of BLG and investigate their interplay with topological gapless states. Vacancies in pristine graphene lead to the appearance of localized states and magnetic moments [29-31]; these states appear at zero energy and are spin split into a couple of spin-polarized states due to the uncompensated sublattices [32-34]. For vacancy defects near the SDW, the defect states interact with the topologically protected states in the gap, and the landscape of the states spanning the gap changes. Despite the strong perturbation caused by vacancies, we find a topologically protected gapless state that is spin split.

The ability to obtain two gapless channels with different spin orientations, i.e., a characteristic of topological insulators, is valuable. Although the spin is not locked to the momentum, we find that one of the two spin channels is strongly locked to the layer with vacancies. This result is our most significant observation and implies that vacancy defects can drive the current flow in the domain wall predominantly to one layer and one spin orientation, which indicates a potential route for the design of a new type of electronic device that could exploit features combining spintronics and layertronics [25].

## II. MODEL AND METHODS

We investigate BLG with vacancies close to a SDW, as schematically shown in Fig. 1. We note that other experimental realizations of domain walls have common physics parameters, which can be described in the first instance by considering the minimal wall [18, 21, 23, 24]. The system studied is divided into three regions [35]: pristine BLG on the left- and right-hand sides of the SDW is treated as the leads and the central region includes the SDW with nearby vacancies located in the upper layer. There is translation symmetry along the SDW, with the unit cell shown in Fig 1 (b). To calculate the local density of states (LDOS), we employ the Green function matching technique [35-38]. The Hamiltonians for the inner region and the leads, as well as the Green function for the entire system are self-consistently calculated in the mean-field Hubbard model within the π-electron approximation. [39] The value of the on-site Coulomb interaction term is set to U = 2.8 eV [40, 41]. The Green function and the resulting LDOS are k-dependent due to the periodicity.

## III. RESULTS AND DISCUSSION

We now study the electronic structure of gated BLG with a SDW decorated with vacancies, as shown in Fig. 1. A voltage V = 0.18 eV is applied to the bottom layer. The corresponding LDOS for the middle region including the SDW with vacancies, as per Fig. 1, is shown in Fig. 2. The green and yellow arrows identify the spin-up and spin-down bands, respectively. The spin-up and spin-down states in the valence and conduction band continua are superimposed because the induced spin splitting is considered to be almost negligible.

Away from the cone on the left- and right-hand sides, we find two spin-polarized states. These stem from the spin-split vacancy states, which strongly interact with the corresponding topological states close to the cone due to the SDW. A detailed analysis of these vacancy states confirms their strong localization in the upper layer; the spin-polarized LDOS in the upper layer is a few orders of magnitude higher than that in the lower layer.

At the cone, two spin-polarized states span the entire energy gap. They appear as a result of the interaction and mixing of the two spin-degenerate topological states at the SDW with the spin-split vacancy band. The most striking result is the persistence of a topological spin-split state in the gap, despite the strong interaction of the electronic structure of BLG with the SDW caused by the vacancy defects.

We also perform calculations for other systems with different distances between consecutive vacancies or between vacancies and the SDW. No significant differences are observed in the trends of the results presented in Fig. 2. The spin-splitting of the gapless state decreases when the distance between vacancies increases. Similarly, when the vacancies move away from the SDW region, their role in the extended gapless states weakens, and the spin splitting slightly decreases. The decrease in the spin-splitting of the topological state for the sparse distribution of vacancies can be easily explained; the gapless states are spatially located along the SDW and decay away from the domain wall [18].

The total magnetic moment in the middle region is slightly below 0.6 $\mu_B$. The moment obtained by uncompensated sublattices due to the vacancy is reduced from the a priori value of 1 $\mu_B$ because the leads pass charge to the vacancy and the SDW region. Furthermore, the induced magnetic moment in the topologically protected state is smaller than the vacancy moment, and the moments are antiferromagnetically coupled.

To understand the interplay between the topologically protected states and vacancy states, we remove the observed spin polarization. We performed LDOS calculations with the on-site Coulomb interaction set to zero, i.e., hindering spin-splitting of the involved states. In the first instance, we find one gapless state before spin polarization is included. This finding is further clarified by modelling the interaction of the spin-degenerate gapless states with the localized vacancy state. We use a three-band Hamiltonian:

$$H_{\text{vac-SDW}}(k) = \begin{pmatrix} -\alpha k & \frac{1}{2}\gamma_1 & \gamma_2 \\ \frac{1}{2}\gamma_1 & -\alpha k + V & 0 \\ \gamma_2 & 0 & 0 \end{pmatrix}, \quad (1)$$

where $\gamma_1$ is the interlayer hopping, $V$ is the gate voltage responsible for the gap opening, and $\gamma_2$ represents the interaction between the gapless states and the vacancy state at zero energy. The diagonal $-\alpha k$ terms are related to the topological states at the K valley; the zero-energy diagonal term denotes the flat vacancy state. We include the vacancies in the top layer and the gate in the bottom layer. By solving the eigenproblem for $H_{\text{vac-SDW}}$ it becomes clear that the interaction between the topological states and the vacancy state implies the survival of a topological state along the gap. Other results and further discussion are presented in the Supplementary Information.

We next investigate the spin-polarization gain of the topological states after including the vacancy states and how this gain depends on the gate voltage. We consider the layer-resolved LDOS for the gapless state at the Fermi energy. This LDOS is calculated by accounting for the contribution from the atoms in the middle region and integrating over $k$-values in a range of $\pm 0.1$ around the point where the topological state crosses the Fermi level. Fig. 3 (a) and (b) show the LDOS in the lower and upper layers for two values of gate voltage $V = 0.18$ eV and $V = 0.36$ eV, respectively. The LDOS of topological states is mainly spin-down and localized strongly in the upper layer, especially because the vacancy is in the same layer. The magnetic coupling shows opposite spin polarization to the vacancy state, as noted above. The separation of the spin-up and spin-down gapless states depending on $k$ values, different for voltages $V < \gamma_1$ and $V > \gamma_1$, is shown in Figs. 3 (a) and (b), respectively. This reflects the fact that with an increasing voltage, the calculated spin splitting decreases because the energy of the π-vacancy state decreases for higher voltages and interacts more with the valence bulk continuum. The spin-down contribution of the topological state increases and is primarily localized in the upper layer, especially for voltages $V > \gamma_1$.

The spin polarization of the topological state for experimentally attainable voltages is mainly spin down, and thus almost all the current flow in the SDW will have the same spin orientation mostly localized in one layer, as shown schematically in Fig. 3 (c). Therefore, the presence of vacancies in BLG with a SDW affords not only spin splitting of gapless states but also allows for partial locking of the spin to a single layer. This effect could allow the exploitation of BLG in spintronics and relates to the role of magnetic impurities in general.

## IV. CONCLUSIONS

We searched for a method to spin split the topologically protected gapless states, which are induced by SDW, that appear in gated BLG to render them useful in spintronic applications. Spin splitting can be accomplished by introducing magnetic defects, which we modelled as π-vacancies. We found that despite the strong perturbation of the topological gapless states caused by the presence of vacancies, a topological state connecting the valence and conduction bands survived in the gap. Its spin degeneracy was lifted, and the state was locked in one graphene layer. We thus obtained two spin-polarized channels for current conductance along the domain wall, which is similar to topological insulators. Although the focus of our study was on vacancies, the results could also be extrapolated to magnetic atoms near a SDW.

### Data Availability

The data that support the fndings of this study are available from the corresponding author upon reasonable request.

### Competing Interests

The authors declare no competing interests.

### Author Contributions

W. J. conceived the project and performed all numerical calculations. A. A. introduced the model Hamiltonian to illustrate the results. The manuscript was written through contributions of both authors. Both authors have given approval to the final version of the manuscript.

Acknowledgments

This work was partially supported by Project FIS2016-76617-P of the Spanish Ministry of Economy and Competitiveness MINECO, the Basque Government under the ELKARTEK project (SUPER), and the University of the Basque Country (Grant No. IT-756-13). AA and WJ acknowledge the hospitality of the Institute of Physics at Nicolaus Copernicus University and the Donostia International Physics Center, respectively.

[1] K. S. Novoselov, A. K. Geim, S. V. Morozov, D. Jiang, Y. Zang, S. V. Dubonos, I. V. Grigorieva, and A. A. Firsov, Science **306**, 666 (2004).

[2] K. S. Novoselov, A. K. Geim, S. V. Morozov, D. Jiang, M. I. Katsnelson, I. V. Grigorieva, S. V. Dubonos, and A. A. Firsov, Nature **438**, 197 (2005).

[3] Y. Cao, Y. Cao, V. Fatemi, S. Fang, K. Watanabe, T. Taniguchi, E. Kaxiras, and P. Jarillo-Herrero, Nature **556**, 43 (2018).

[4] Y.-M. Lin and P. Avouris, Nano Letters **8**, 2119 (2008).

[5] F. Schwierz, Nature Nanotechnology **5**, 487 (2010).

[6] S.-M. Choi, S.-H. Jhi, and Y.-W. Son, Nano Letters **10**, 3486 (2010).

[7] F. Xia, D. B. Farmer, Y. Lin, and P. Avouris, Nano Lett. **10**, 715 (2010).

[8] J. E. Padilha, M. P. Lima, A. J. R. da Silva, and A. Fazzio, Phys. Rev. B **84**, 113412 (2011).

[9] H. Santos, A. Ayuela, L. Chico, and E. Artacho, Phys. Rev. B **85**, 245430 (2012).

[10] Q. Zhang, Y. Yaofeng, K. S. Chan, Z. Mu, and J. Li, Applied Physics Express **1**, 075104 (2018).

[11] T. Ohta, A. Bostwick, T. Seyller, K. Horn, and E. Rotenberg, Science **313**, 951 (2006).

[12] E. V. Castro, K. S. Novoselov, S. V. Morozov, N. M. R. Peres, J. M. B. L. dos Santos, J. Nilsson, F. Guinea, A. K. Geim, and A. H. C. Neto, Phys. Rev. Lett. **99**, 216802 (2007). [13] Y. Zhang, T.-T. Tang, C. Girit, Z. Hao, M. C. Martin, A. Zettl, M. F. Crommie, Y. R. Shen, and F. Wang, Nature (London) **459**, 820 (2009).

[14] B. N. Szafranek, D. Schall, M. Otto, D. Neumaier, and H. Kurz, Applied Physics Letters **96**, 112103 (2010).

[15] J. S. Alden, A. W. Tsen, P. Y. Huang, R. Hovden, L. Brown, J. Park, D. A. Muller, and P. L.

McEuen, Proc. Natl. Acad. Sci. **110**, 11256 (2013).

[16] A. Vaezi, Y. Liang, D. H. Ngai, L. Yang, and E.-A. Kim, Phys. Rev. X **3**, 021018 (2013).

[17] P. San-Jose, R. V. Gorbachev, A. K. Geim, K. S. Novoselov, and F. Guinea, Nano Lett. **14**, 2052 (2014).

[18] M. Pelc, W. Jaskolski, A. Ayuela, and L. Chico, Phys. Rev. B **92**, 085433 (2015).

[19] P. Maher, C. R. Dean, A. F. Young, T. Taniguchi, K. Watanabe, K. L. Shepard, J. Hone, and P. Kim, Nature Phys **9**, 154 (2013).

[20] F. Zhang, A. H. MacDonald, and E. J. Mele, Proc. Natl. Acad. Sci. U.S.A. **110**, 10546 (2013).

[21] L. Ju, Z. Shi, N. Nair, Y. Lv, C. Jin, J. Velasco Jr, C. Ojeda-Aristizabal, H. A. Bechtel, M. C. Martin, A. Zettl, et al., Nature **520**, 650 (2015).

[22] J. Li, K. Wang, K. J. McFaul, Z. Zern, Y. Ren, K. Watanabe, T. Taniguchi, Z. Qiao, and J. Zhu, Nat Nano **11**, 1060 (2016).

[23] J. Lin, W. Fang, W. Zhou, A. R. Lupini, J. C. Idrobo, J. Kong, S. J. Pennycook, and S. T. Pantelides, Nano Lett. **13**, 3262 (2013).

[24] W. Jaskólski, M. Pelc, L. Chico, and A. Ayuela, Nanoscale **8**, 6079 (2016).

[25] W. Jaskólski, M. Pelc, W. G. Bryant, L. Chico, and A. Ayuela, 2D Materials **5**, 025006 (2018).

[26] E. H. Lieb, Phys. Rev. Lett. **62**, 1201 (1989).

[27] E. J. G. Santos, D. Sanchez-Portal, and A. Ayuela, Phys. Rev. B **81**, 125433 (2010).

[28] E. J. G. Santos, A. Ayuela, and D. Sanchez-Portal, New J Phys **14**, 043022 (2012).

[29] V. M. Pereira, F. Guinea, J. M. B. Lopes dos Santos, N. M. R. Peres, and A. H. Castro Neto, Phys. Rev. Lett. **96**, 036801 (2006).

[30] J. J. Palacios, J. Fernandez-Rossier, and L. Brey, Phys. Rev. B **77**, 195428 (2008).

[31] M. P. Lopez-Sancho, F. de Juan, and M. A. H. Vozmediano, Phys. Rev. B **79**, 075413 (2009).

[32] K. Kusakabe, Carbon-based magnetism (Elsevier, Amsterdam, 2006).

[33] W. Jaskólski, A. Ayuela, M. Pelc, H. Santos, and L. Chico, Phys. Rev. B **83**, 235424 (2011).

[34] S. Dutta and K. Wakabayashi, Sci. Rep. **5**, 11744 (2015).

[35] M. B. Nardelli, Phys. Rev. B **60**, 7828 (1999).

[36] S. Datta, Electronic Transport in Mesoscopic Systems (Cambridge University Press, Cambridge, 1995).

[37] L. Chico, L. X. Benedict, S. G. Louie, and M. L. Cohen, Phys. Rev. B **54**, 2600 (1996).

[38] W. Jaskólski and L. Chico, Phys. Rev. B **71**, 155405 (2005).

[39] The intralayer hopping $\gamma_0 = -2.7$ eV and a single interlayer hopping $\gamma_1 = 0.1\gamma_0$ [11, 12] parameters are used.

[40] D. Gunlycke, D. A. Areshkin, J. Li, J. W. Mintmire, and C. T. White, Nano Letters **7**, 3608 (2007).

[41] W. Jaskolski, L. Chico, and A. Ayuela, Phys. Rev. B **91**, 165427 (2015).

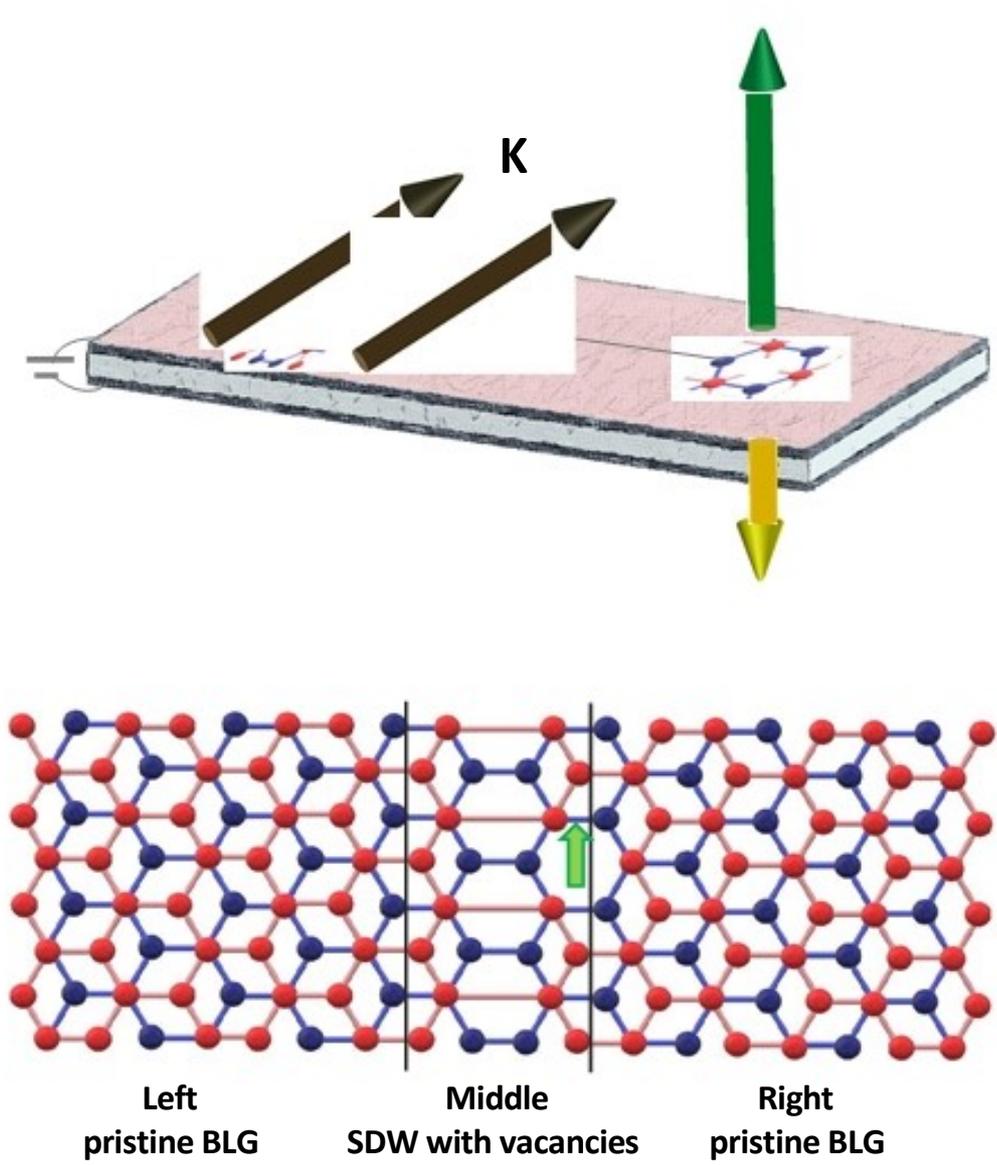

FIG. 1: (a) Schematic representation of the investigated BLG, including a SDW and vacancies. The topological gapless states locked to the K valley and momentum are displayed as a couple of arrows along the SDW. The green-yellow arrows indicate the vacancy and the related up-down components in the magnetic moment.(b) An example of the studied unit cells. The SDW with vacancies is situated in the upper (red) layer and embedded between the left and right leads of pristine BLG.

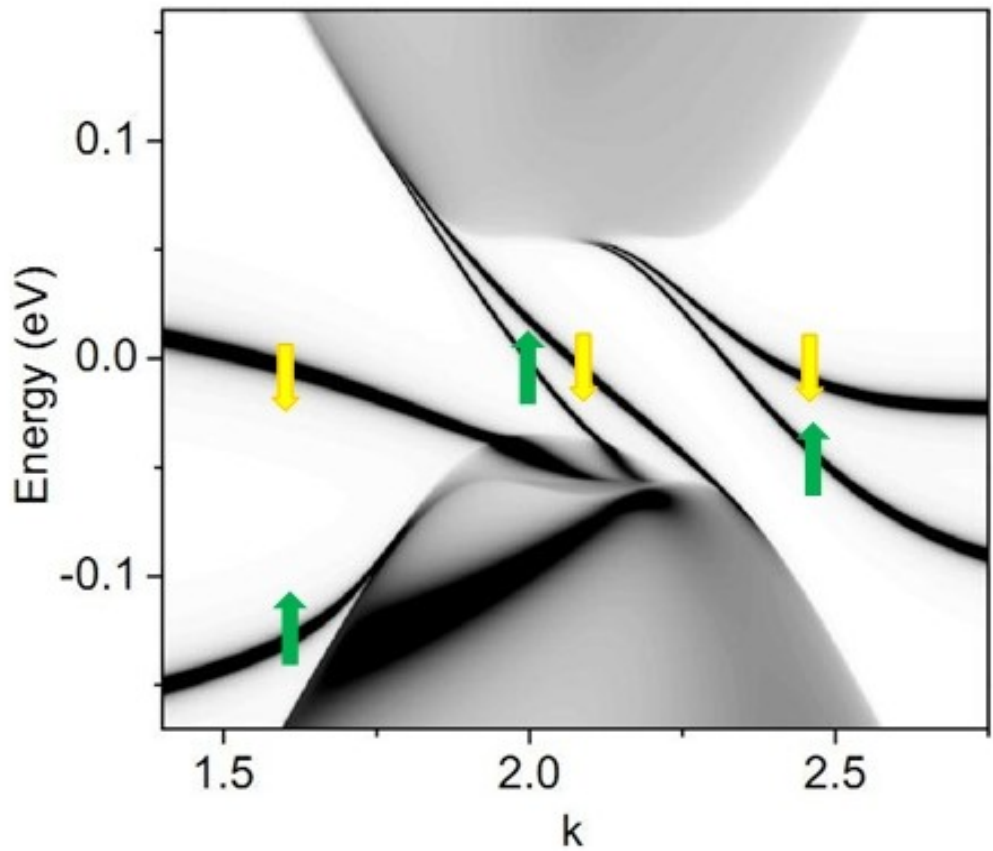

FIG. 2: Energy structure near the Fermi level (zero-energy) for gated BLG with SDW and vacancies. The k values are given in units of π/a, where a is the period length in the zigzag direction and set as a = 1. A gapless state is present in the band gap connecting the valence and conduction bands at the cone, and it is spin split due to the interaction with the vacancy band.

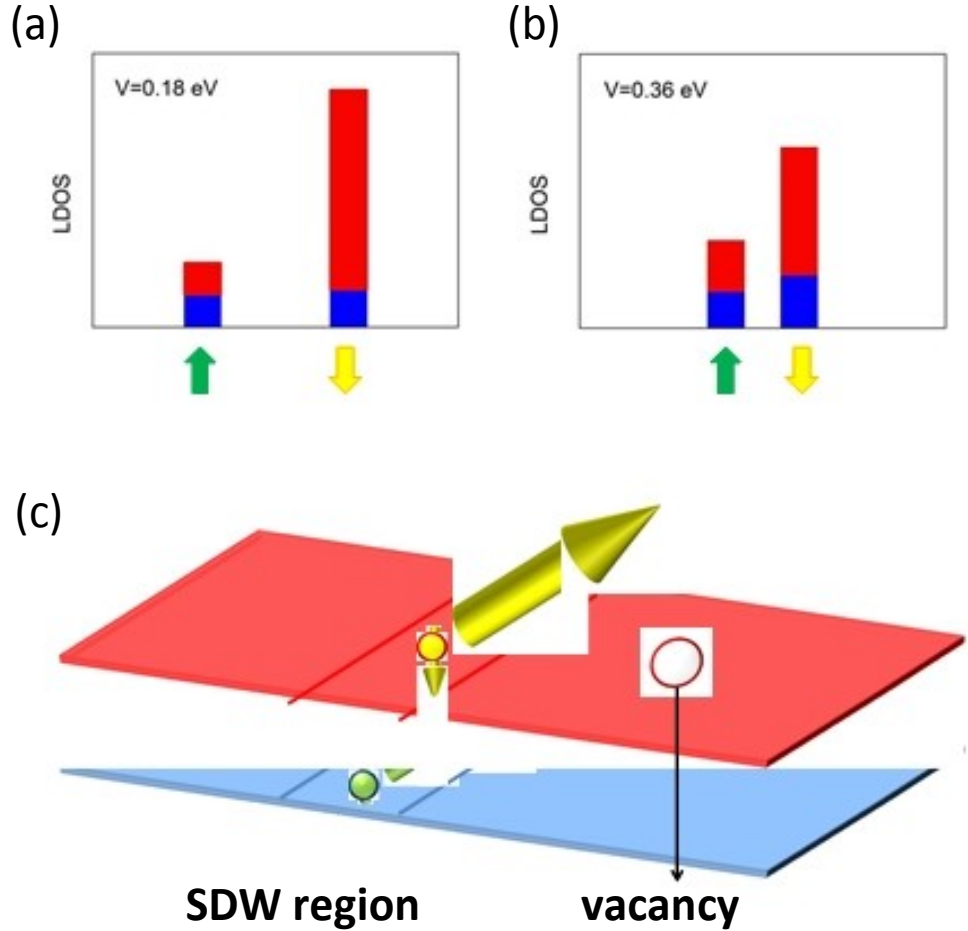

FIG. 3: (a) and (b) Layer-resolved LDOS of spin-polarized gapless states shown in Fig. 2 (marked by green and yellow arrows) for the case of BLG with SDW and vacancies for two different gate voltages. Layers are denoted by colours: blue - bottom, red - top. LDOS values are calculated at the Fermi energy and are represented by bar heights. (c) Visualization of current flow along the SDW in BLG with vacancies. The sizes of the dots and arrows represent the current and spin densities, respectively, in the layers. A current with almost fixed spin orientation flows mainly in the upper layer.